\documentclass[a4paper]{article}
\usepackage{graphicx}

\begin{document}

\noindent
{\bf SAGA-HE-151-99}

\bigskip

\begin{center}
{\LARGE \bf{Nonlinear effects in Schwinger-Dyson Equation}}
\end{center}

\centerline{Hiroaki Kouno$^*$, Akira Hasegawa, Masahiro Nakano$^{**}$ and Kunito Tuchitani}

\centerline{Department of Physics, Saga University Saga 840, Japan}

\centerline{**University of Occupational and Environmental Health, Kitakyushu 807, Japan}

\bigskip

\centerline{*e-mail address:kounoh@cc.saga-u.ac.jp}


\bigskip

\bigskip

\centerline{\large\bf{Abstract}}

We study nonlinear effects in the QED ladder Schwinger-Dyson(SD) equation. 
Without further approximations, 
we show that all nonlinear effects in the ladder SD equation can be included in the effective couplings and how a linear approximation works well. 
The analyses is generalized in the case of the improved ladder calculation with the Higashijima-Miransky approximation. 

\bigskip

\bigskip


The dynamical symmetry breaking (DSB) is widely studied by the Schwinger-Dyson (SD) equation. 
(See references \cite{rf:Higashijima1} and \cite{rf:Miransky1}, and the references therein. )
In QED, the lowest-order approximation is a ladder approximation. 
It was pointed out that there is DSB of the chiral symmetry in the QED ladder SD equation with a finite cut off. 
(See, e.g.,  \cite{rf:Maskawa,rf:Fukuda,rf:Miransky1}. ) 
In QCD, the calculation was modified by introducing a running coupling. 
The modified calculation is called an improved ladder approximation and gives a good results even quantitatively. 

Although the original SD equation is an integral equation, it is sometimes rewritten in the form of a nonlinear differential equation. (See, e.g., \cite{rf:Fukuda,rf:Higashijima3,rf:Higashijima1,rf:Miransky1}. ) 
Even in the most simple ladder approximation, the nonlinear equation can not be solved exactly and a linear approximation is widely used. (See, e.g., \cite{rf:Fomin,rf:Miransky1}. ) 
The aim of this paper is to re-examine the nonlinear effects in the QED ladder SD equation and show how the linear approximation works well. 
After that, we generalize our analyses in the case of the improved ladder calculation with the Higashijima-Miransky approximation. \cite{rf:Higashijima2,rf:Miransky2}


In Landau gauge, the QED ladder Schwinger-Dyson equation is given by 
\begin{equation}
B(p^2)=3e^2\int{d^4q\over{(2\pi)^4}}{B(q^2)\over{q^2+B^2(q^2)}}{1\over{(p-q)^2}}+m_0,
\label{eq:1}
\end{equation}
where $B(p^2)$ is a fermion self-energy with four momentum $p$ and $m_0$ is a bare mass of the fermion. 
We restrict our discussions on Euclidean region only. 
Putting $p=me^{t\over{2}}$($m$ is a some energy scale) and rescaling $x(t)=B(p^2)/p$, after some algebra, we get 
a nonlinear second differential equation \cite{rf:Fukuda,rf:Higashijima3,rf:Higashijima1}
\begin{equation}
{\ddot{x}}+2{\dot{x}}+{3\over{4}}x+\lambda{x\over{1+x^2}}=0
\label{eq:2}
\end{equation}
with two boundary conditions 
\begin{equation}
v_1\equiv e^{3t\over{2}}({\dot{x}}+{1\over{2}}x)\rightarrow 0 \qquad (t \rightarrow -\infty) 
\label{eq:3}
\end{equation}
and 
\begin{equation}
v_3\equiv e^{t\over{2}}({\dot{x}}+{3\over{2}}x)\rightarrow {m_0\over{m}} \qquad (t \rightarrow  \infty).  
\label{eq:4}
\end{equation}
In these equations, $\lambda ={3e^2\over{16\pi^2}}$, $\dot{x}={dx\over{dt}}$ and $\ddot{x}={d^2x\over{dt^2}}$. 
Since we are interested in DSB of chiral symmetry, we put $m_0=0$. 
The ultra-violet boundary condition (UVBC) (\ref{eq:4}) becomes
\begin{equation}
v_3\equiv e^{t\over{2}}({\dot{x}}+{3\over{2}}x)\rightarrow 0 \qquad (t \rightarrow  \infty).  
\label{eq:ad19a}
\end{equation}
The equation (\ref{eq:2}) can be rewritten as
\begin{equation}
{dv_1\over{dt}}=-\lambda {xe^{{3t\over{2}}}\over{1+x^2}}
\quad {\rm or} \quad
{dv_3\over{dt}}=-\lambda {xe^{{t\over{2}}}\over{1+x^2}}.
\label{eq:ad17}
\end{equation}

We can construct a potential $V(x)$ as \cite{rf:Cohen}
\begin{equation}
V(x)={3\over{8}}x^2+{\lambda\over{2}}\log{(1+x^2)}; \qquad {dV(x)\over{dx}}={3\over{4}}x+\lambda{x\over{1+x^2}}
\label{eq:5}
\end{equation}
and a hamiltonian $H(y,x)$
\begin{equation}
H(y,x)={1\over{2}}y^2+V(x),
\label{eq:6}
\end{equation}
where $y={\dot{x}}$ is a conjugate momentum of $x$. 
This system is a dissipative system and it is easy to derive 
\begin{equation}
{dH\over{dt}}=-2y^2\leq 0
\label{eq:7}
\end{equation}
The equality in (\ref{eq:7}) is hold only $y=0$. 
Because $y=0$ and $x\neq 0$ is not a stable point, $H$ decreases. 
Since the potential $V(x)$ has a minimum only at $x=0$, Eq. (\ref{eq:7}) gives
\begin{equation} 
H(y,x), \quad V(x) \rightarrow 0 
\label{eq:8}
\end{equation}
in the limit of $t\rightarrow \infty$. 
Next, we define 
\begin{equation}
r\equiv \sqrt{x^2+y^2}
\label{eq:9}
\end{equation}
In the large $t$ limit, $r^2$ also approaches zero, since
\begin{equation}
{3\over{8}}r^2\leq H(y,x) \leq \max\left( {1\over{2}}, {3\over{8}}+{\lambda\over{2}} \right)r^2. 
\label{eq:10}
\end{equation}
Therefore, $x$ and $y$ approach zero in this limit. 
Similarly, we can show 
\begin{equation}
r^2 \rightarrow \infty, 
\label{eq:11}
\end{equation}
when $t$ approaches $-\infty$. 
Since $x=0$ and $y=\dot{x}\neq \pm \infty$ is not a stable point, 
we consider the case $x^2\rightarrow \infty $ in the limit $t\rightarrow -\infty$.  
In this case, the nonlinear equion (\ref{eq:2}) is reduced to a linear equation 
\begin{equation}
{\ddot{x}}+2{\dot{x}}+{3\over{4}}x=0. 
\label{eq:12}
\end{equation}
This equation can be solved analytically and we get \cite{rf:Fukuda}
\begin{equation}
x(t)=C_1 e^{-{t\over{2}}}+C_3e^{-{3t\over{2}}},
\label{eq:13}
\end{equation}
where $C_1$ and $C_3$ are some constants. 
To satisfy the infra-red boundary condition (IRBC) (\ref{eq:3}), $C_3=0$. 
We remark that, due to Eq. (\ref{eq:ad17}), $v_1$ and $v_3$ are conserved in this limit. 
On the other hand, in the high-energy region in which $x\rightarrow 0$, the nonlinear equation (\ref{eq:2}) reduces to another linear equation as 
\begin{equation}
{\ddot{x}}+2{\dot{x}}+{3\over{4}}x+\lambda x=0
\label{eq:14}
\end{equation}
The general solution in this limit is given by \cite{rf:Fukuda} 
\begin{eqnarray}
x(t)&=&c_1 e^{-(1-a)t}+c_2e^{-(1+a)t} \qquad (\lambda <1/4)
\nonumber
\\
x(t)&=&c_3 e^{-t}+c_4te^{-t} \qquad (\lambda =1/4)
\nonumber
\\
x(t)&=&c_5 e^{-t}\cos{(bt +c_6)} \qquad (\lambda >1/4), 
\label{eq:15}
\end{eqnarray}
where $a =\sqrt{1/4-\lambda}$, $b = \sqrt{\lambda -1/4}$ and $c_1\sim c_6$ are some constants. 
Therefore, in the high-energy region, the solution has oscillating behavior in the case of $\lambda >\lambda_c\equiv 1/4$ and it can connect the line $v_1=0$ ($y=-x/2$) of IRBC and the line $v_3=0$ ($y=-3x/2$) of UVBC with $m_0=0$. 
In that case, the equation (\ref{eq:1}) has a non-trivial solution $B(p^2)\neq 0$ and DSB of chiral symmetry takes place. \cite{rf:Maskawa,rf:Fukuda,rf:Miransky1} 
Therefore, $\alpha_c\equiv 4\pi\lambda_c/3=\pi /3$ is called a critical coupling. 
Below, for simplicity, we call $\lambda_c$ itself a "critical coupling". 
 
The condition $\lambda >\lambda_c$ is necessary but is not sufficient for DSB. 
If the solution continues to oscillate in the ultraviolet region, it may go beyond the line $y=-3x/2$ of UVBC. 
On the other hand, the integral (\ref{eq:1}) may diverge. 
To converge the integral, a form factor (or a cutoff) which switches off the interaction at very large $t$ may be introduced. 
Due to the form factor, the equation (\ref{eq:2}) becomes (\ref{eq:12}) again and the solution ceases to oscillate. 
If the solution satisfies UVBC when the interaction is switched off by the form factor, it evolutes on the line $y=-3x/2$ after that time. 
This is a sufficient condition for DSB. 
The aim of this paper is to re-examine the nonlinear effects in this scenario and show how the linear approximation works well. 
To achieve this, we define 
\begin{equation}
x=r\cos{\theta}\quad{\rm and}\quad y=r\sin{\theta }.
\label{eq:16}
\end{equation}
We get the differential equations in this polar coordinate. 
\begin{equation}
{dr\over{dt}}=\sin{\theta}\left( -2r\sin{\theta}+{1\over{4}}r\cos{\theta}-\lambda {r\cos{\theta}\over{1+r^2\cos^2{\theta}}}\right).
\label{eq:17}
\end{equation}
\begin{equation}
{d\theta\over{dt}}=-1-2\cos{\theta}\sin{\theta}+{\cos^2{\theta}\over{4}}-\lambda {\cos^2{\theta}\over{1+r^2\cos^2{\theta}}}.
\label{eq:18}
\end{equation}
The equation (\ref{eq:18}) can be rewritten as 
\begin{equation}
{d\theta\over{dt}}=-{y^2+2xy+(3/4+\lambda_{eff}(x) )x^2\over{r^2}}=-{y^2+2xy+\omega^2(x)x^2\over{r^2}}, 
\label{eq:19}
\end{equation}
where $\lambda_{eff}(x)=\lambda/(1+x^2)(\leq \lambda)$ is an effective coupling and $\omega^2(x)={dV(x)\over{dx}}/x=3/4+\lambda_{eff}(x)$. 
The ${d\theta\over{dt}}$ becomes negative unless a second order equation for $y$
\begin{equation}
y^2+2xy+(3/4+\lambda_{eff}(x) )x^2=0
\label{eq:20}
\end{equation}
has real solutions. 
This condition gives
\begin{equation}
\lambda_{eff} (x) \leq 1/4=\lambda_c.
\label{eq:21}
\end{equation}
In this case, the real solutions are given by 
\begin{equation}
y=-x\pm \sqrt{x^2(1-\omega^2(x))}=-x\pm\sqrt{x^2\left({1\over{4}}-\lambda_{eff}(x)\right)}. 
\label{eq:22}
\end{equation}
These are boundaries of a region where ${d\theta\over{dt}}> 0$. 
We call this region "typhoon-like" because the solution has a $left$-handed rotation in this region. 
On the boundaries ${d\theta \over{dt}}=0$. 
For convenience, we include the boundaries in the typhoon-like region. 
In the remaining part of $x$-$y$ plane, ${d\theta \over{dt}}<0$. 
We call this region "anti-typhoon-like" because the solution has a $right$-handed rotation in this region. 
It is interesting that two boundaries lie between the line $y=-x/2$ of IRBC and the line $y=-3x/2$ of UVBC on $x$-$y$ plane. 
In particular, two boundaries coincide with the line $y=-x/2$ and the line $y=-3x/2$ when $\lambda=0$. 
In fig. 1, we show the boundaries in the cases of $\lambda <\lambda_c$ and $\lambda >\lambda_c$. 
The typhoon-like region is continuous at the origin of $x$-$y$ plane, if $\lambda \leq \lambda_c$. 
It is discontinuous at the origin, if $\lambda >\lambda_c$. 

Our picture of DSB is as following. 
In the infra-red limit, the solution is given by $x=C_1e^{-t/2}$. 
As $t$ becomes large, the interaction is switched on and the conservation of $v_1$ is violated, i.e., the solution deviates from $x=C_1e^{-t/2}$. 
As $x$ becomes smaller, the effective coupling $\lambda_{eff}(x)$ becomes larger. 
However, if $\lambda\leq \lambda_c$, the effective coupling $\lambda_{eff}(x)$ can not be larger than $\lambda_c$. 
Therefore, there is a typhoon-like region just between the line $y=-x/2$ of IRBC and the line $y=-3x/2$ of UVBC. 
Since the solution has a left-handed rotation within the typhoon-like region, the solution is prevented from connecting the line $y=-x/2$ with the line $y=-3x/2$ in the same quadrant. 
On the other hand, the anti-typhoon-like region prevents the solution connecting the line $y=-x/2$ with the line $y=-3x/2$ in the different quadrant. 
If $\lambda >\lambda_c$, at some $t$, $\lambda_{eff}(x)$ becomes greater than $\lambda_c$. 
The typhoon-like region disappears. 
The solution begins to oscillate and can connect IRBC and UVBC. 
The DSB may happen. 
Therefore, $\lambda >\lambda_c$ is a necessary condition for DSB. 

As is seen above, the nonlinear effects can be included in the effective coupling $\lambda_{eff}(x)$. 
The typhoon-like region also prevents the solution deviating much from the line $y=-x/2$ of IRBC. 
Therefore, $x=C_1e^{-t/2}$ may be a good approximation for the exact solution, until $\lambda_{eff}(x)$ becomes $\lambda_c$. 
Furthermore, if $\lambda$ is not much larger than $\lambda_c$, $x$ becomes sufficiently small before $\lambda_{eff}(x)$ becomes $\lambda_c$, and, therefore, the linear approximation (\ref{eq:14}) works well in the region where $\lambda_{eff}(x)>\lambda_c$. 
In this case, the effective coupling $\lambda_{eff}(x)$ can be approximated as $\lambda/(1+(C_1)^2e^{-t})$. 
This is nothing but the linear approximation used by Fomin, Gusynin and Miransky. \cite{rf:Fomin,rf:Miransky1} 

Next, we consider a sufficient condition for DSB. 
Suppose that $\lambda$ is slightly larger than $\lambda_c$ and $x$ does not deviate much from $y=\dot{x}=-x/2$ until some $t_0$ where $r$ becomes sufficiently small. 
We can re-define $t_0$ as an origin of $t$. 
Neglecting the $r$-dependence, 
we integrate the equation (\ref{eq:18}) with an initial conditions $\dot{x}(0)=-x(0)/2$ and impose a final condition $\dot{x}(t_{cut})=-3x(t_{cut})/2$ where $t_{cut}$ is a cutoff parameter. 
The result gives the condition 
\begin{equation}
b t_{cut}=n\pi-\theta^* \qquad (n=1,2...),
\label{eq:23}
\end{equation}
where $b$ is the same as in Eq. (\ref{eq:15}) and $\theta^* =\arctan{{4b\over{1-4b^2}}}$. 
This is an approximate version of the condition for DSB in \cite{rf:Miransky1}. 

Our analyses can be generalized in more general equation as 
\begin{equation}
{\ddot{x}}+2\gamma(t,x){\dot{x}}+\omega^2(t,x)x
={\ddot{x}}+2\gamma(t,x){\dot{x}}+{\partial V(t,x)\over{\partial x}}=0. 
\label{eq:24}
\end{equation}
In this case, we get 
\begin{equation}
{dH\over{dt}}=-2\gamma (t,x)y^2+{\partial V(t,x)\over{\partial t}}
\label{eq:25}
\end{equation}
\begin{equation}
{d\theta\over{dt}}=-{y^2+2\gamma(t,x)xy+\omega^2(t,x)x^2\over{r^2}} 
\label{eq:26}
\end{equation}
As is already seen, if we put $\gamma (t,x)=1$ and $\omega(x,t)=3/4+\lambda/(1+(C_1)^2e^{-t})$ in Eq. (\ref{eq:24}), we get the linear approximation used by Fomin, Gusynin and Miransky. \cite{rf:Fomin,rf:Miransky1} 
If we put $\gamma (t,x)=1$ and $\omega(x,t)=3/4+\lambda\Theta (t)$ where $\Theta (t)$ is a step function, it reproduces the results in Eq. (\ref{eq:23}). 

To apply our analyses to the improved ladder calculation, consider the integral
\begin{equation}
B(z)={1\over{z}}\int_0^zdu{\lambda(z,u)B(u)u\over{u+B^2(u)}}
+\int_z^\infty  du{\lambda(z,u)B(u)\over{u+B^2(u)}}+m_0, 
\label{eq:29}
\end{equation}
where $z=p^2=m^2e^t$ and $u=q^2$. 
For $\lambda (x,y)$, we use the Higashijima-Miransky approximation \cite{rf:Higashijima2,rf:Miransky2} 
\begin{equation}
\lambda(z,u)=\lambda(z)\Theta(z-u)+\lambda(u)\Theta (u-z),
\label{eq:31}
\end{equation}
Differentiating $B(z)$ twice with respect to $z$, we get
 the differential equation as Eq. (\ref{eq:24}) with \cite{rf:Higashijima3,rf:Higashijima1}
\begin{equation}
\gamma(t,x)=\gamma(t)=1-{1\over{2}}U(t);\quad U(t)\equiv {\dot{\lambda}^*(t)\over{\lambda^*(t)}}
\label{eq:32}
\end{equation}
and  
\begin{equation}
\omega^2 (t,x)={1\over{x}}{\partial V(t,x)\over{\partial x}};\quad 
V(t,x)\equiv {1\over{2}}\left( {3\over{4}}-{1\over{2}}U(t) \right) x^2+{\lambda^*(t)\over{2}}\log{(1+x^2)}, 
\label{eq:33}
\end{equation}
where $\lambda^*(t)=\lambda (t)-\dot{\lambda}(t)$. 
The IRBC and UVBC are modified as 
\begin{equation}
v_1^\prime \equiv e^{3t\over{2}}{({\dot{x}}+{1\over{2}}x)\over{\lambda^*(t)}}\rightarrow 0 \qquad (t \rightarrow -\infty) 
\label{eq:34}
\end{equation}
and 
\begin{equation}
v_3^\prime \equiv e^{t\over{2}}\left({\lambda (t)\over{\lambda^*(t)}}{\dot{x}}+\left(1+{1\over{2}}{\lambda (t)\over{\lambda^*(t)}}\right)x\right)\rightarrow {m_0\over{m}} \qquad (t \rightarrow  \infty).  
\label{eq:35}
\end{equation}
We remark that IRBC (\ref{eq:34}) is reduced to the original one $v_1=0$ if $\lambda^*(t)$ is finite in the limit $t\rightarrow -\infty$. 
Due to (\ref{eq:26}), if 
\begin{equation}
D=x^2(\gamma^2(t)-\omega^2(t,x))=x^2\left({1\over{4}}(1-U(t))^2-\lambda^*_{eff}(t,x)\right);\quad \lambda^*_{eff}(t,x)\equiv {\lambda^*(t)\over{1+x^2}} 
\label{eq:36}
\end{equation}
becomes negative, a typhoon-like region disappears. 
The condition (\ref{eq:36}) gives $\lambda_{eff}^*(t,x)> 1/4(1-U(t))^2$. 
In the case of $D\geq 0$, the boundaries of the typhoon-like region is given by 
\begin{equation}
y=-x\gamma(t)\pm \sqrt{D}. 
\label{eq:37}
\end{equation}
If $\dot{\lambda}(t)$ and $\ddot{\lambda}(t)$ is small compared with $\lambda (t)$, we get $\gamma (t)\approx 1$, $\omega^2 (x,t)\approx 3/4+\lambda(t)/(1+x^2)$ and $v_3^\prime \approx v_3$. 
In this case, approximately, $\max (\lambda (t)) > \lambda_c$ is a necessary condition for DSB.


In summary, we have re-examined the solution of DSB of chiral symmetry in the ladder SD equation. 
Without further approximation, we have showed that the nonlinear effects are included in the effective couplings $\lambda_{eff}(x)$. 
If $\lambda_{eff}(x)>\lambda_c$,  the typhoon-like region disappears and DSB takes place. 
The typhoon-like region prevents the solution deviating much from IRBC and 
makes the linear approximation works well. 
The analyses is generalized in the case of the improved ladder calculation with the Higashijima-Miransky approximation.


\begin{center}
{\large\bf{Acknowledgements}}
\end{center}

The authors thank K-I. Aoki and M. Maruyama for useful discussions. 
One of the author (H.K.) also thank O. Kiriyama and F. Takagi for useful discussions. 

\vspace{2cm}


\begin{figure}
\begin{center}
    \includegraphics*[height=6cm]{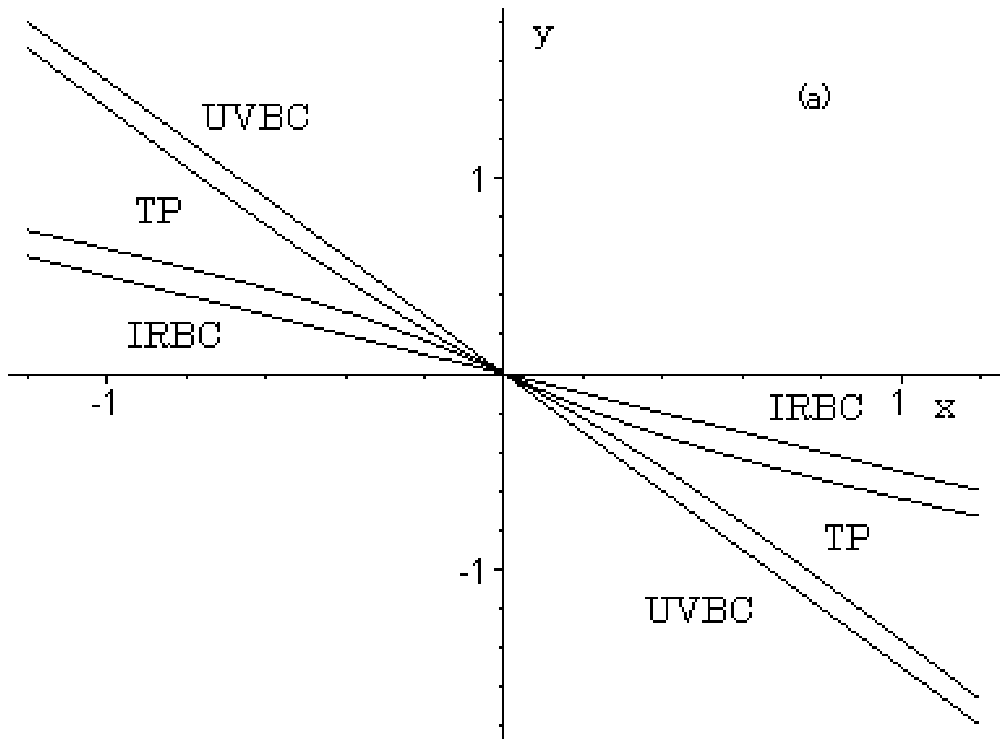}
    \includegraphics*[height=6cm]{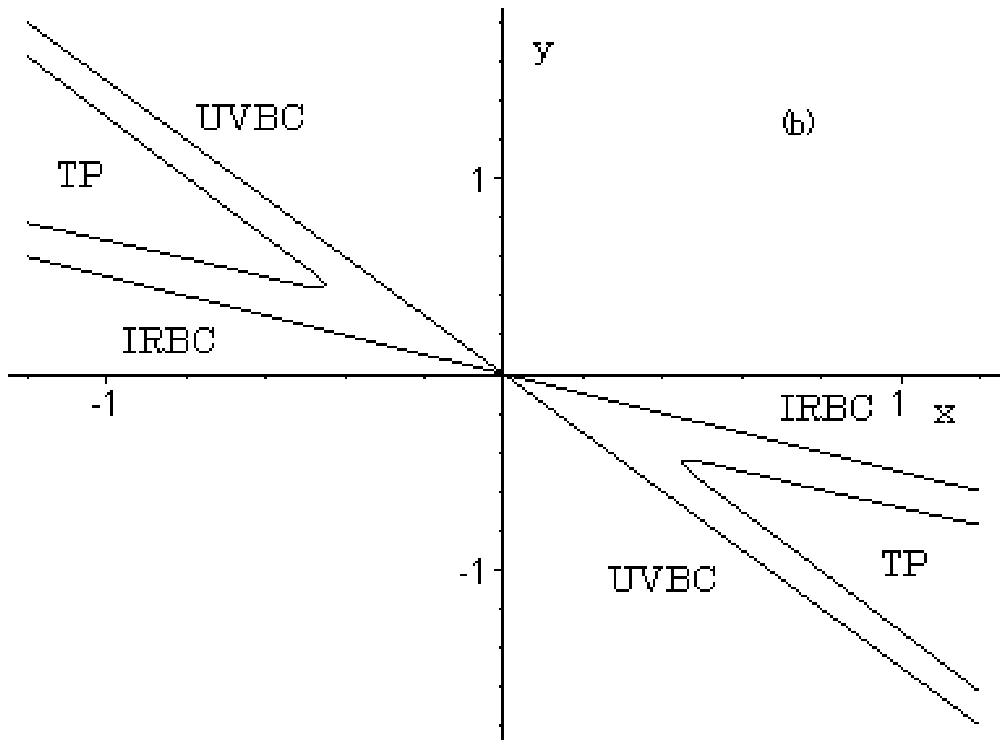}
\caption{The typhoon-like region (denoted by "TP") and its boundaries in $x$-$y$ plane. The remaining part in the plane is anti-typhoon-like. Two straight lines are IRBC and UVBC with $m_0=0$. (a) $\lambda <\lambda_c$. (b) $\lambda >\lambda_c$. }
\label{fig1}
\end{center}
\end{figure}
\end{document}